\newcommand{\non}{\nonumber}
\newcommand{\cov}[1]{\mbox{/\hspace{-1.9mm}#1}}
\documentclass[showpacs,preprintnumbers,amsmath,amssymb]{revtex4}
\usepackage{graphicx}
\usepackage{dcolumn}
\usepackage{bm}
\begin{document}
\title{\bf Effective Meson Field Theory from QCD}
\author{Ron-Chou Hsieh}
\email{hsieh@hepth.phys.ncku.edu.tw} \affiliation{Department of
Physics, National Cheng Kung University\\ Tainan, Taiwan 701, ROC}
\date{\today}
\begin{abstract}
We give a simple and straightforward procedure of how to construct
an effective meson Lagrangian from QCD Lagrangian. We integrate
the methods of Gasser, Leutwyler, Alkofer and Reinhardt and use
the derivative expansion scheme to derive the low energy effective
Lagrangian for meson fields to $O(p^4)$. In this paper, why the
meson particle can be treated as the goldstone mode is very clear.
In our calculation the result in $O(p^2)$ is the same as in the
chiral perturbation theory, but the result in $O(p^4)$ is
different from that in literature. We will discuss the
discrepancies and give some remarks on our result.
\end{abstract}
\maketitle
\section{Introduction}  \label{Sec.1}
In the late fifties it was found that the perturbative method of
quantum field theory always fails when it was applied to the
strong interactions of hadrons (especially for the light meson
systems). Although the remarkable success in QED, it was
recognized that perhaps we should have a suitable formulation
which is independent of perturbative approach to solve this. In
fact the scaling behavior of the structure functions in the deep
inelastic region tells us that the perturbative method can only be
implemented in the high energy region. Now we know this is because
QCD is a non-Abelian Yang-Mills theory with gauge group ${\rm
SU}_{\rm c}(3)$. Due to the property of asymptotic freedom in this
theory, we can safely use perturbative method to handle high
energy reaction processes. But in the low energy region, the
coupling constant is too large to use perturbative expansion. So
if we can find an effective Lagrangian with new degrees of freedom
(DOFs) and enough small coupling constant in the low energy
region, then we still can use perturbative method. The question is
how it can be done.

Free quarks are not observed in nature. It is commonly assumed
that quarks are confined (color confinement), {\it i.e.}, the DOFs
can only be hadrons. If we want to find an effective Lagrangian
which describes adequate low energy behavior of particle physics,
the quark DOF must be replaced by the hadron DOF. In 1984, Gasser
and Leutwyler~\cite{Gasser:84a,Gasser:85a} proposed a method about
how to systematically extract the low-energy structure of strong
interactions from experiments. They considered only light
pseudoscalar particles and used the spontaneously chiral symmetry
breaking property in the low energy region. It is an effective
chiral Lagrangian with the Nambu-Goldstone-bosons (NGB) DOF only.
The form of each term in the chiral Lagrangian is determined by
chiral symmetry requirements.

The theoretic derivative method also have been investigated. In
the literature~\cite{Espriu:90a,Ebert:86a} we can see how the
procedure is implemented. A critical step is to introduce some NGB
depended terms (by symmetry consideration) into the fermionic
determinant and make a coordinate-space expansion of it. In
general, the proper-time regularization scheme has been used. By
this technique the meson DOF can be extracted and the effective
Lagrangian with only meson DOF are obtained. Although the physical
results obtained by the effective Lagrangian which derived from
these methods indeed coincide with the experiment well. We still
have two questions need to overcome. How the effective Lagrangian
can be derived from the first principle and why the meson particle
can be treated as the goldstone mode. In this article we will try
to solve this question and derive an effective meson Lagrangian
from QCD Lagrangian. We shall use a different method, derivative
expansion method~\cite{Dobado:97a}, which is a general expansion
scheme used in effective field theory, to derive the low-energy
effective Lagrangian. Using this method, we find that the loop
effect in the underlying theory can be represented naturally in
all types of terms in effective Lagrangian and the obtained low
energy coefficients (LECs) have been renormalized. However, some
of our results in $O(p^4)$ are different from that in
others'~\cite{Ebert:86a,Espriu:90a,Gasser:84a,Gasser:85a}. Latter
we will discuss the discrepancies and give some remarks on our
results.

In this paper we first give a brief introduction to the effective
field theory and illustrate how to use the derivative expansion
method to obtain an effective Lagrangian from its underlying
theory. These will be done in Sec.~\ref{Sec.2}. Afterward we
briefly review some results derived by Alkofer and
Reinhardt~\cite{Alkofer:95a}, who obtained an effective QFD
Lagrangian by integrating out the gluon DOF in the QCD Lagrangian.
We then mesonize this Lagrangian and rewrite it as a constituent
quark model (CQM) Lagrangian. This procedure will be illustrated
in Sect.~\ref{Sec.3}. In this section we can see how the
quark-gluon dynamics are represented by the constituent quarks
propagating in background meson fields. Furthermore, we can find
that as the running energy going down the potential of background
meson fields will have spontaneously symmetry breaking which is
essential condition of linear sigma model. The obtained Lagrangian
is similar to that in the nonlinear sigma model except that the
meson fields are background fields. Because we want to have an
effective Lagrangian with only the meson DOFs. The quark fields
should be integrated out. In Sec.~\ref{Sec.4} we will illustrate
how to use the derivative expansion method introduced in
Sec.~\ref{Sec.2} to obtain what we want and discuss the difference
between our result and others'. There we also give some remarks on
our result. Finally, our conclusion is given in Sec.~\ref{Sec.5}.

\section{Derivative expansion method} \label{Sec.2}
The aim of an effective field theory is to find an adequate and
effective Lagrangian to describe the physical systems of interest.
For example, owing to the property of asymptotic freedom the QCD
Lagrangian can only be used in high energy region. So if we want
to cope with the low-energy physical problems we must find an
effective low-energy Lagrangian to describe it. Since effective
field theories should be based on the fundamental field theory.
They must be the quantum field theoretical implementation of the
quantum ladder of the fundamental theory. As the energy changes,
the new DOFs become relevant and must be included in the theory.
Then we should ask: how these new DOFs are related to the old DOFs
and how these new DOFs can be extracted from the underlying theory
and made into an effective theory of its own. In this section we
shall focus on how to extract the new DOFs from the underlying
theory. Below we shall use a beautiful method, derivative
expansion method, which mentioned in~\cite{Dobado:97a} to do that.
Firstly, we assume that the relations between the new and old DOFs
have been known and the corresponding effective Lagrangian with
the new and old DOFs have been obtained. Then the next step is to
derive the effective Lagrangian with only new DOFs. Consider a
simple model with the Lagrangian
\begin{equation}
{\cal L}(\phi,\Phi) = \frac{1}{2}\partial_\mu \phi\partial^\mu
\phi - \frac{1}{2}m^2{\phi}^2 + \frac{1}{2}\partial_\mu
\Phi\partial^\mu \Phi - \frac{1}{2}M^2{\Phi}^2 +
\frac{\alpha}{2}{\phi}^2{\Phi}^2,
\end{equation}
whose action is
\begin{equation}
S[\phi,\Phi]=\int d^4x{\cal
L}(\phi,\Phi)\,\equiv\,S[\phi]+S_{\Phi}[\phi,\Phi],
\end{equation}
where
\begin{equation}
S[\phi]=\int d^4x\left(\frac{1}{2}\partial_\mu \phi\partial^\mu
\phi - \frac{1}{2}m^2{\phi}^2\right),
\end{equation}
\begin{equation}
S[\phi,\Phi]=\int d^4x\left(\frac{1}{2}\partial_\mu
\Phi\partial^\mu \Phi - \frac{1}{2}M^2{\Phi}^2 +
\frac{\alpha}{2}{\phi}^2{\Phi}^2\right),
\end{equation}
and its generating functional is
\begin{equation}
Z=\int{\cal D}\phi{\cal D}\Phi e^{iS[\phi,\Phi]} = \int{\cal
D}\phi e^{i\Gamma_{eff}[\phi]},
\end{equation}
where the $\phi$ and $\Phi$ indicate new and old DOFs
respectively. Here the normalization constant was abbreviated by
simplification and if no special indications we will keep this
simplification in the following equations. Because the $\phi$
field is the DOF we want to keep, we integrated out the $\Phi$
field. Then we have
\begin{equation}
e^{i\Gamma_{\rm eff}[\phi]} = \int{\cal D}\Phi e^{iS[\phi,\Phi]} =
e^{iS[\phi]}\int{\cal D}\Phi e^{iS_{\Phi}[\phi,\Phi]} =
e^{iS[\phi]}(\det \hat{O})^{-1/2},
\end{equation}
\begin{equation}
\hat{O}_{xy}=(-\Box_x-M^2+\alpha\phi^2_x)\delta_{xy},
\end{equation}
\begin{equation}
\Gamma_{\rm eff}[\phi]=S[\phi]+\frac{i}{2}{\rm Tr}\ln\hat{O},
\end{equation}
and now we can write
\begin{equation} \label{trexpand}
{\rm Tr}\ln \hat{O} = {\rm Tr}\ln(-\Box-M^2) + {\rm
Tr}\ln\left[1-\alpha(\Box+M^2)^{-1}\phi^2\right].
\end{equation}
We can ignore the first term in eq.(\ref{trexpand}) because it
does not depend on $\phi$ field and can be absorbed into the
normalization constant. After expanding the logarithm term, we
obtain
\begin{equation}
\Gamma_{\rm eff} = S[\phi] +
\frac{i}{2}\sum^{\infty}_{n=1}\frac{(-1)^n}{n}
Tr[\alpha(\Box+M^2)^{-1}\phi^2]^n \equiv S[\phi] +
\sum^{\infty}_{n=1}\Gamma^{(n)}[\phi].
\end{equation}
The operator $(\Box+M^2)^{-1}$ is the propagator for the free
$\Phi$ field and can be defined as
\begin{equation}
\Delta_{xy} \equiv \langle x|(\Box+M^2)^{-1}|y\rangle = \int
\frac{d^4q}{(2\pi)^4} e^{-iq(x-y)}\frac{1}{q^2-M^2+i\epsilon}.
\end{equation}
Hence all the terms involving $\Phi$ were integrated out and we
can find that each term in the effective action $\Gamma_{eff}$ is
composed of $\phi$ field and the propagator of $\Phi$ field. In
other words, the loop contribution of $\Phi$ field has been
absorbed into the new terms in effective Lagrangian. For example,
the $\Gamma^{(1)}[\phi]$ contribution is
\begin{equation}
\Gamma^{(1)}[\phi]=-\frac{i}{2}\alpha\int d^4x\Delta_{xx}\phi^2_x,
\end{equation}
which is corresponds to a one-loop effect with internal $\Phi$
field line.

In the following sections we shall use this method to expand the
fermionic determinant. At first we introduce how to obtain a
constituent quark form Lagrangian from QCD.

\section{Constituent quark form} \label{Sec.3}
In QCD with three light flavors, u, d and s, the generating
functional for Green functions is defined via the path-integral
formula
\begin{equation} \label{eq:2-1}
Z=\int {\cal D}q{\cal D} \overline{q} {\cal D}A_{\mu}
\exp\left\{i\int d^{4}x{\cal L}^0_{\rm QCD}\right\},
\end{equation}
where the ${\cal L}^0_{\rm QCD}$ is QCD Lagrangian in the absence
of quark masses,
\begin{equation} \label{QCD-Lagrangian}
{\cal L}^0_{\rm QCD} = -\frac{1}{4}G^a_{\mu \nu}G_a^{\mu \nu} +
i\overline{q}_L \gamma^{\mu}D_{\mu}q_L + i\overline{q}_R
\gamma^{\mu}D_{\mu}q_R,
\end{equation}
$G^a_{\mu\nu}$ is the antisymmetric tensor of the gluon field.
This Lagrangian is not adequate for perturbative calculations in
the low energy region. Our aim is to find an effective Lagrangian
${\cal L}_{\rm eff}$ to describe the low energy physics and make
the perturbative method work safely. We have known that in the low
energy region the effective DOFs are hadron fields. So the first
step is to extract the hadron DOFs from the QCD Lagrangian. For
this purpose, we rewrite the generating functional such as
\begin{equation} \label{g-integrated}
Z = \int{\cal D}q{\cal D} \overline{q} \exp\left[i\int
d^{4}x\overline{q}(i\gamma^{\mu}\partial_{\mu})q +
i\Gamma[j]\right],
\end{equation}
where
\begin{equation}
\Gamma[j] = -i\ln\int{\cal D}A_{\mu}\exp\left[i\int d^4x
\left(-\frac{1}{4}G^a_{\mu \nu}G_a^{\mu\nu} + g
A^a_{\mu}j^{\mu}_a\right)\right].
\end{equation}
The effective action of the gluon field $\Gamma[j]$ can be
expanded in power of the quark current $j^a_{\mu}$ as
\begin{equation}  \label{g-action}
\Gamma[j] = \Gamma[j=0] + \sum^{\infty}_{n=1}\frac{1}{n!}\int
\left\{\Gamma^{(n)}(x_1,x_2\cdots,x_n)^{a_1\cdots a_n}
_{\mu_1\cdots\mu_n}\prod^n_{i=1}
\left[j^{\mu_i}_{a_i}(x_i)dx_i\right]\right\},
\end{equation}
where the expansion coefficients are
\begin{equation}
\Gamma^{(n)}(x_1,x_2\cdots,x_n)^{a_1\cdots a_n}
_{\mu_1\cdots\mu_n} = \left[\frac{\delta^{(n)}\Gamma[j]}
{\prod^n_{i=1}\delta j_{a_i}^{\mu_i}(x_i)}\right]_{j=0} =(-i)\cdot
(ig)^n\left<0|\mathrm{T}\left[A^{a_1}_{\mu_1}(x_1)A^{a_2}_{\mu_2}(x_2)\cdots
A^{a_n}_{\mu_n}(x_n)\right]|0\right>,
\end{equation}
and the brackets $\left<\cdots\right>$ are the irreducible n-point
correlation functions of the gluon field. Now let's consider the
second order term of the effective action of the gluon field,
$\Gamma_2[j]$ (the 0th and 1st order term have no contribution).
Using the Fierz decomposition~\cite{Greiner:96a} of spin, color,
flavor matrices and by taking the Feynman gauge condition
\begin{equation} \label{Apropagator}
\left<A^a_{\mu}(x)A^b_{\nu}(y)\right> = -
i\delta^{ab}g_{\mu\nu}\int\frac{d^4k}{(2\pi)^4}
\frac{e^{-ik\cdot(x-y)}}{k^2-m_g^2+i\epsilon} =
-i\delta^{ab}g_{\mu\nu}\Delta(x-y),
\end{equation}
where the effective gluon mass $m_g$ originated in higher order
correctional effect~\cite{Greiner:94a}. Then the $\Gamma_2[j]$
is~\cite{Alkofer:95a,Cahill:85a}\footnotemark[1]
\begin{eqnarray} \label{2nd-g-action}
\Gamma_2[j] &=& \frac{g^2}{2}\int\int d^4x d^4y
j^a_{\mu}(x)\left[\int\frac{d^4k}{(2\pi)^4}\frac{e^{-ik
\cdot(x-y)}}{k^2-m_g^2+i\epsilon}\right]j^{\mu a}(y)\non\\
& = & \frac{g^2}{3}\int\int d^4x d^4y
\overline{q}(y)\Lambda_\alpha q(x)\Delta(x-y)
\overline{q}(x)\Lambda^\alpha q(y),
\end{eqnarray}
\footnotetext[1]{Here the original result has meson and diquark
parts. Since we only want to discuss the meson system, we ignore
the diquark part.} where the matrix $\Lambda^\alpha$ is tensor
product of color, flavor and Dirac matrices
\begin{equation} \label{meson matrix}
\Lambda_\alpha=\mbox{1\hspace{-1.3mm}1}_C\otimes(\frac{\lambda^A}{2})_F
\otimes\Gamma_{\alpha},\qquad A=0, \ldots 8
\end{equation}
and
\begin{equation}
\Gamma_{\alpha}\in\left\{\mbox{1\hspace{-1.3mm}1},i\gamma_5,
\frac{i}{\sqrt{2}}\gamma_{\mu},\frac{i}{\sqrt{2}}\gamma_{\mu}\gamma_5\right\}.
\end{equation}
Applying the eqs.(\ref{g-action}) and (\ref{2nd-g-action}) in
eq.(\ref{g-integrated}) one obtains
\begin{equation}
Z = \int{\cal D}q{\cal D} \overline{q} \exp\left\{i\int
d^{4}x\overline{q}(i\gamma^{\mu}\partial_{\mu}) q -
\frac{1}{2}\int\int d^4x d^4y \overline{q}(y)\Lambda_\alpha
q(x)\Delta'(x-y) \overline{q}(x)\Lambda^\alpha
q(y)+\cdots\right\},
 \label{eq:2-4}
\end{equation}
with
\begin{equation}
\Delta'(x-y)=\frac{2}{3}g^2\Delta(x-y) = -i\cdot
\frac{g^2}{6\pi^2}\cdot\frac{m_g}{l}K_{-1}(m_g l),
\end{equation}
where $l=\sqrt{(x^0-y^0)^2+(\vec{x}-\vec{y})^2}$ and $K_{-1}(m_g
l)$ is modified Bessel function. This is a NJL-like Lagrangian. It
can be mesonized by introducing an auxiliary bilocal field for
which $\eta(x,y)=\eta_\alpha(x,y) \Lambda^\alpha$ and
$\eta^*(x,y)=\eta(y,x)$ via the identity
\begin{eqnarray}
\exp\left[-\frac{1}{2}\int\int d^4x d^4y\overline{q}(y)
\Lambda_\alpha q(x)\Delta'(x-y)\overline{q}(x)\Lambda^\alpha
q(y)\right]\hspace{5cm}\non\\ \hspace{3cm} = \frac{ (\det
\Delta')^{-1/2}}{(2\pi)^2} \int{\cal D}\eta\exp\left\{-\int\int
d^4x d^4y \left[\frac{1}{2}\eta^*_\alpha \Delta'^{-1}\eta^\alpha -
i\overline{q}(y)\Lambda^\alpha q(x)\eta_\alpha\right]\right\}
 \label{eq:2-6}
\end{eqnarray}
The generating functional then becomes
\begin{equation}
Z = \int{\cal D}q{\cal D}\overline{q}{\cal D}\eta \exp\left\{i\int
d^{4}x\overline{q}(i\gamma^{\mu}\partial_{\mu}) q - \int\int d^4x
d^4y \left[\frac{1}{2}\eta^*_\alpha \Delta'^{-1}\eta^\alpha
 - i\overline{q}(y)\Lambda^\alpha q(x)\eta_\alpha\right]+ \cdots\right\}.
 \label{bilocal}
\end{equation}
In the eq.(\ref{bilocal}) we can find that the strong interaction
effect in the original Lagrangian eq.(\ref{QCD-Lagrangian}) has
been replaced by the background meson field $\eta$. Furthermore,
the vacuum expectation value of the auxiliary field is
\begin{equation}
\left<\eta_\alpha(x,y)\right> = i\Delta'(x-y)\left<\overline{q}(x)
\Lambda_\alpha q(y)\right>.
\end{equation}
This indeed demonstrate that the meson ($\eta$ field) is composed
of quark and anti-quark fields. Because the bilocal property of
the $\eta$ field should complicate the problem, we will try to
localize it. We know that the DOFs in the low energy region are
meson particles. So we can make an assumption that for small
momentum the positions of the meson's constituent particles,
$\overline{q}(x)$ and $q(y)$, are undifferentiated and by
inserting the Dirac delta function
$\delta(x^0-y^0)\cdot\delta^3(\bar{l}-|\vec{x}-\vec{y}|)$ into the
generating functional eq.(\ref{bilocal}) to constraint it, where
$l_0$ is the space between the quarks and $\bar{l}$ approaches to
zero. Then we have
\begin{eqnarray}
\int\int d^4x d^4y\delta(x^0 - y^0)\cdot\delta^3(\bar{l} -
|\vec{x}-\vec{y}|)
\cdot\left[\frac{1}{2}\eta^*_\alpha\Delta'^{-1}\eta^\alpha
- i\overline{q}(y)\Lambda^\alpha q(x)\eta_\alpha\right]\non\\
\approx-i\int d^4x\left[\frac{1}{g_r^2}\eta^*_\alpha\eta^\alpha +
\overline{q}\Lambda^\alpha q\eta_\alpha\right],\hspace{3cm}
\end{eqnarray}
where we have absorbed all other coefficients into the
renormalized coupling constant $g_r$. After localizing the bilocal
term we can further decompose the generic meson field $\eta$
according to its properties under Lorentz transformation
\begin{equation}
\eta=S+i\gamma_5P-i\mbox{/\hspace{-1.9mm}V}-\mbox{/\hspace{-1.9mm}A}\gamma_5,
 \label{eq:2-8}
\end{equation}
where $S$ is a scalar, $P$ is a pseudoscalar, $V$ is a vector and
$A$ is a axialvector field. All these fields are flavor matrices,
$S = S^{\alpha}\left(\lambda_{\alpha}/\sqrt{2}\right)_F$, etc.
Then we can discuss the behavior of all kinds of meson multiplets.
For simply we only consider scalar and pseudoscalar particles. We
choose $V_{\mu} = A_{\mu} = 0$, $S =
g_{\phi}\sigma\mbox{1\hspace{-1.3mm}1}_F/\sqrt{N}$, $P =
g_{\phi}\vec{\lambda}\cdot\vec{\phi}/\sqrt{2}$, with $N$ is its
flavor number and $g_{\phi}$ is Yukawa coupling constant, then the
effective Lagrangian in eq.(\ref{bilocal}) is
\begin{equation} \label{LS-type}
{\cal L} = i\overline{q}\cov{$\partial$}q +
g_{\phi}\left(\sigma\overline{q}q/\sqrt{N} +
i\vec{\phi}\cdot\overline{q}\gamma_5\vec{\lambda}q/\sqrt{2}\right)
-V(\sigma^2+\vec{\phi}^2).
\end{equation}
And the potential $V(\sigma^2+\vec{\phi}^2)$ is\footnotemark[2]
\footnotetext[2]{We have known that the $\eta$ field can be
treated as the composite field of quark and anti-quark. And the
$\eta^2$ term can be derived from the $j^2$ contribution, so we
expect that the $\eta^4$ term also can be derived from the $j^4$
contribution and the coefficient $\lambda$ should be proportional
to $\frac{g_{\phi}^4}{g_r^4}$.}
\begin{equation} \label{ssbpotential}
V(\sigma^2+\vec{\phi}^2) = -\frac{g_{\phi}^2}{2g_r^2}
(\sigma^2+\vec{\phi}^2) +
\frac{\lambda}{4!}(\sigma^2+\vec{\phi}^2)^2.
\end{equation}
Here we only take the quadratic and quartic terms into account
because that in the low energy region the expansion coefficient is
so small so that there is no need to consider higher order terms.
Besides, we also discard the cubic
term\footnotemark[3]\footnotetext[3]{In fact, this term is an
explicit chiral symmetry breaking term.}, for we still don't know
how to deal with it. Now let's look at the form of this potential.
Because the strong coupling constant $g_r$ is in denominator and
it is a running energy depended parameter. For the high energy
region the expanding coefficient will be too large to implement
the perturbative expansion. But as the running energy going down
there will be spontaneous chiral symmetry breaking emerged from
this potential and then the pseudoscalar particles can be treated
as its goldstone modes. In order to see this the ground state
should be decided firstly. For the stable vacuum we have the
constraint condition
\begin{equation} \label{f-definition}
\sigma^2+\vec{\phi^2}=\frac{6g_{\phi}^2}{g_r^2\lambda} = Nf^2
\end{equation}
with $f$ is a running-energy depended parameter which will be
defined latter. The constraint can be resolved by choosing
\begin{eqnarray} \label{constraint}
\sigma(x) & = & \sqrt{N}f\cos\left(
\frac{\sqrt{2}|\vec{\phi}(x)|}{\sqrt{N}f}\right),\non\\
\vec{\phi}(x) & = & \sqrt{N}f\hat{\phi}\sin\left(
\frac{\sqrt{2}|\vec{\phi}(x)|}{\sqrt{N}f}\right)
\end{eqnarray}
with $\vec{\phi}(x)=|\vec{\phi}(x)|\hat{\phi}$ and
$\hat{\phi}\cdot\hat{\phi}=1$. Substituting eq.(\ref{constraint})
into the eq.(\ref{LS-type}) we find that the Lagrangian can be
rewrote as
\begin{eqnarray} \label{CQM-L}
{\cal L} &=& i\overline{q}\cov{$\partial$}q +
g_{\phi}f\overline{q}\left[\cos\left(
\frac{\sqrt{2}|\vec{\phi}(x)|}{\sqrt{N}f}\right) +
i\gamma_5\frac{\vec{\lambda}\cdot\hat{\phi}\sqrt{N}}{\sqrt{2}}\sin\left(
\frac{\sqrt{2}|\vec{\phi}(x)|}{\sqrt{N}f}\right)\right]q\non\\
& = & \overline{\psi}(i\cov{$\partial$} - \cov{$\overline{V}$} -
\cov{$\overline{A}$}\gamma_5 + m_{\psi})\psi \,\equiv\, {\cal
L}_{\rm CQM}
\end{eqnarray}
with
\begin{equation}
\cov{$\overline{V}$}=-\frac{i}{2}(\xi^\dag\cov{$\partial$}\xi +
\xi\cov{$\partial$}\xi^\dag),\hspace{4mm}\cov{$\overline{A}$}=
-\frac{i}{2}(\xi^\dag\cov{$\partial$}\xi -
\xi\cov{$\partial$}\xi^\dag), \hspace{4mm}\psi=\Lambda
q,\hspace{4mm}\overline{\psi}=\overline{q}\Lambda,
\hspace{4mm}m_{\psi}=g_{\phi}f,
 \label{eq:2-10}
\end{equation}
\begin{equation}
\Lambda = \exp{\left\{i\sqrt{2}\gamma_5\Phi/(2f)\right\}} =
P_R\xi+P_L\xi^{\dagger},\hspace{6mm} \xi =
\exp{\left\{i\sqrt{2}\Phi/(2f)\right\}},
 \label{eq:2-11}
\end{equation}
and the $\Phi$ is the conventional parametrization matrix of meson
field:
\begin{equation}
\Phi(x) \equiv \frac{\vec{\lambda}\cdot\vec{\phi}}{\sqrt 2} =
\left[
\begin{array}{ccc}
\frac{1}{\sqrt 2}\pi^0 + \frac{1}{\sqrt 6}\eta & \pi^+ & K^+ \cr
\pi^- & - \frac{1}{\sqrt 2}\pi^0 + \frac{1}{\sqrt 6}\eta & K^0 \cr
K^- & \bar K^0 & - \frac{2}{\sqrt 6}\eta
\end{array}
\right].
  \label{eq:phi_matrix}
\end{equation}
The Lagrangian in eq.(\ref{CQM-L}), which reveals constituent
quarks propagating in the background meson fields, can be regarded
as describing a constituent quark model. It is noted that the sign
of the constituent quark's mass term is positive. It means that in
our model the constituent quark should have negative mass.
Although this can be adjusted by change the sign of the parameter
$g_\phi$. Later we will find that if we want to obtain an
effective meson Lagrangian with positive mass, the constituent
quark's mass must be negative. This is different from the result
of Espriu, de Rafael and Taron~\cite{Espriu:90a}.

So far we obtain an effective Lagrangian with the new DOFs,
constituent quark fields $\psi$ and meson fields $\vec{\phi}$.
Although we start with a massless QCD Lagrangian, the constituent
quark fields are massive. We can treat this as a remnant effect of
the strong interactions after integrating out the gauge fields.

\section{Effective meson Lagrangian} \label{Sec.4}
We shall now attempt to derive an effective Lagrangian with meson
DOFs only. In the latest section we have obtained an effective CQM
Lagrangian. We can find that the form is the same as in the
nonlinear sigma model except that the meson fields are background
fields. In fact we can already safely use the perturbative
expansion method in this effective CQM Lagrangian to handle the
low energy problems. But this is not our aim here. Our purpose is
to transform the QCD theory into an effective theory with that the
DOFs are completely in the meson form. So let's get down to
business. Firstly, let us add some external current source terms
to the massless QCD Lagrangian eq.(\ref{QCD-Lagrangian}). Then the
corresponding constituent quark form Lagrangian in
eq.(\ref{CQM-L}) can be rewritten such this
\begin{eqnarray} \label{addsource}
{\cal L}_{\rm CQM}^{\rm (new)} & = & {\cal L}_{\rm CQM} +
\overline{q}{\gamma}^{\mu}(v_{\mu} + {\gamma}_5 a_{\mu})q -
\overline{q}(s-i{\gamma}_5 p)q \non \\
& = &  \overline{\psi}(i\mbox{/\hspace{-1.9mm}$D'$} +
m_{\psi})\psi,
\end{eqnarray}
with
\begin{eqnarray}
i\cov{$D$}' & = & i\cov{$\partial$}+\cov{$G$}' \non \\
& = & i\cov{$\partial$} - \cov{$\overline{V}$} -
\cov{$\overline{A}$}\gamma_5 \,+\, \gamma^\mu(P_L\xi
l_\mu\xi^\dagger + P_R\xi^\dagger r_\mu\xi)\non \\ & &
\,-\,\frac{1}{2B_0}(P_R\xi\chi^\dagger\xi + P_L\xi^\dagger
\chi\xi^\dagger)
 \label{eq:(16)3-2}
\end{eqnarray}
\begin{equation}\label{eq:(17)3-3}
l_\mu = v_\mu - a_\mu,\hspace{4mm}r_\mu = v_\mu +
a_\mu,\hspace{4mm}\chi = 2B_0(s + ip),
\end{equation}
where $v_\mu$, $a_\mu$, $s$ and $p$ are the corresponding current
source fields which are hermitian $3\times3$ matrices in flavor
space and $B_0$ is an arbitrary constant. Because when we analyze
low energy physics, the effective realization of quark current in
meson form must be understood. By this elegant technique the
matrix elements of currents can be calculated in a very
straightforward way. We note that it is started from this
constituent quark form Lagrangian, eq.(\ref{addsource}), the
effective chiral Lagrangian in the chiral perturbation theory can
be obtained by the proper-time regularization
method~\cite{Espriu:90a}. But we don't use this method here. Now
by means of the derivative expansion method introduced in
Sec.~\ref{Sec.2}, we can write the generating functional as
\begin{eqnarray}
Z & = & \int {\cal D}q{\cal D}\overline{q} {\cal D}G_{\mu}
\exp^{\left\{i\int d^{4}x[{\cal L}^0_{\rm QCD} \,+\,
\mbox{source term}]\right\}} \non \\
& = & \int {\cal D}\psi{\cal D}\overline{\psi}{\cal D}U
\exp^{\left\{i\int
d^{4}x[\overline{\psi}(i\mbox{/\hspace{-1.9mm}$D'$}
\,+\, m_{\psi})\psi]\right\}}J(U) \non \\
& = & \int {\cal D}U e^{i\Gamma_{\rm eff}[U]}.
 \label{eq:(18)3-4}
\end{eqnarray}
Here $U=\xi\xi$, which is the same as the exponential
parametrization of the NGB fields using in the chiral perturbation
theory (ChPT) and the effective action $\Gamma_{eff}[U]$ is
\begin{eqnarray} \label{effaction}
\Gamma_{\rm eff}[U] & = & \int d^4x{\cal L}_{eff}(U)+\Gamma_{\rm WZW}\non\\
& = & i\sum_{n=1}^\infty\frac{(-1)^n}{n} {\rm
Tr}[(i\cov{$\partial$}+m_\psi)^{-1}\cov{$G$}']^n + \Gamma_{\rm
WZW}\non\\ & \equiv & \sum_{n=1}^\infty\Gamma^{(n)}.
\end{eqnarray}
The Jacobian $J(U)$ comes from the chiral transformation $q \to
\Lambda q$ and $\Gamma_{\rm WZW}$ is the corresponding action,
which is the Wess-Zumino-Witten(WZW)
term~\cite{Wess:71a,Witten:83a}. We neglect this term for
simplicity in this article. By this expansion formula, the
effective Lagrangian can be expanded in the powers of
$\cov{$G'$}$. But the expanded form is not adequate for
discussion. So we rearrange it according to the powers of momentum
such that
\begin{equation}
{\cal L}_{\rm eff}(U)\,=\,{\cal L}_2 \, + \, {\cal L}_4 \,+\,
{\cal L}_6\,+\, \cdots.
 \label{eq:(20)3-6}
\end{equation}
The details of the calculation are given in the appendix. The
result expanding to $O(p^2)$ is
\begin{equation}\label{O(p^2)}
{\cal L}_2 = -i m_{\psi}^2 N_c I_3 \langle D_{\mu}U^{\dagger}
D^{\mu}U \rangle - i(\frac{m_{\psi}}{B_0})N_c I_0 \langle
U^{\dagger}\chi + \chi^{\dagger}U\rangle,
\end{equation}
where $\langle O\rangle$ denotes the trace of the matrix $O$,
$N_c$ is color number and
\begin{equation} \label{covariantD}
D_{\mu}U = \partial_{\mu}U - ir_{\mu}U + iUl_{\mu},\hspace{6mm}
D_{\mu}U^{\dagger} = \partial_{\mu}U^{\dagger} +
iU^{\dagger}r_{\mu} -il_{\mu}U^{\dagger}.
\end{equation}
\begin{equation}\label{I0intergal}
I_0 = \int \frac{d^4q}{q^2-m_{\psi}^2 + i\epsilon} =
\frac{im^2_{\psi}}{(4\pi)^2} \left(N_{\epsilon} + 1 -
\ln\frac{m_{\psi}^2}{{\mu}^2}\right),
\end{equation}
\begin{equation} \label{I3intergal}
I_3 = \int \frac{d^4q}{(q^2-m_{\psi}^2 +
i\epsilon)[(q+p)^2-m_{\psi}^2 + i\epsilon]} =
\frac{I_0}{m^2_{\psi}},
\end{equation}
\begin{equation}
N_{\epsilon} = \frac{2}{\epsilon} + \ln4\pi - \gamma+O(\epsilon),
\end{equation}
where $\gamma$ is Euler's constant. The eq.(\ref{I0intergal}) is a
divergent integral and still need to be renormalized. If we
substituting $m_\psi=g_\phi f$ into the eq.(\ref{I0intergal}) and
renormalized the strong coupling constant again (re-absorbed the
divergent part into the strong coupling constant in the parameter
$f$, eq.(\ref{f-definition})), and let $B_0=m_\psi$,
$g_\phi=\frac{2\pi}{\sqrt{N_c}}$ then we obtain
\begin{equation} \label{renorI}
-iI_0N_c = \frac{(g_\phi f)^2N_c}{(4\pi)^2}= \frac{f^2}{4},
\end{equation}
and
\begin{equation} \label{newO(p^2)}
{\cal L}_2 = \frac{f^2}{4} \langle D_{\mu}U^{\dagger} D^{\mu}U +
U^{\dagger}\chi + \chi^{\dagger}U\rangle.
\end{equation}
We note that the form is totally the same as in
ChPT~\cite{Gasser:85a}. Using the Lagrangian ${\cal L}_2$ the
currents can be calculated by taking the appropriate derivatives
with respect to the external sources:
\begin{subequations} \label{current}
\begin{eqnarray}
J^{\mu i}_L &=& \frac{\delta S_2}{\delta l_{\mu}^i} = -\frac{i
f^2}{2}\langle\lambda^i U\partial^{\mu}U^{\dagger}\rangle
= -f\partial^{\mu}\phi^i+O(\phi^2),\non\\ \\
J^{\mu i}_R &=& \frac{\delta S_2}{\delta r_{\mu}^i} = -\frac{i
f^2}{2}\langle\lambda^i U^{\dagger}\partial^{\mu}U\rangle
= f\partial^{\mu}\phi^i+O(\phi^2).\non\\
\end{eqnarray}
\end{subequations}
Now the physical meaning of the parameter $f$ is obvious. Up to
the order $O(p^2)$, $f$ is equal to the pion decay constant,
$f=f_{\pi^+}\simeq 130.7$ MeV, defined
as\footnotemark[4]\footnotetext[4]{We must emphasized that the
definition, $f=f_{\pi^+}$, was established in the running energy
is low enough. If the running energy is so large so that we can
not ignore the next leading term contribution, then the parameter
$f$ must be redefined.}
\begin{equation}
\langle0|A^-_{\mu}(0)|\pi^+({\bf q})\rangle \,=\,i f_{\pi^+}q_\mu,
 \label{f-definition}
\end{equation}
with
\begin{equation}
A^{\mu i}=\frac{1}{2}(J^{\mu i}_R-J^{\mu i}_L), \hspace{6mm}
A^\pm_{\mu}=\frac{1}{\sqrt{2}}(A^1_\mu\mp i A^2_\mu)
\end{equation}
And the constituent quark mass $m_\psi$ is
\begin{equation}
m_\psi = g_\phi f = \frac{2\pi f}{\sqrt{N_c}}\approx 474.13 \mbox{
Mev}.
\end{equation}
Similarly, the constant $B_0$ can be related to the quark
condensate:
\begin{equation}
\langle0|\overline{q}^i q^j|0\rangle \,=\, -f^2 B_0{\delta}^{ij}
\approx -(200.8\mbox{ Mev})^3{\delta}^{ij}
 \label{eq:(27)3-13}
\end{equation}
If we take $s=M$ and $p=0$, where $M$ denotes the quark-mass
matrices such that
\begin{equation}
M={\rm diag}(m_u,m_d,m_s),
 \label{eq:(28)3-14}
\end{equation}
then the relations between the physical meson masses and the quark
masses are:
\begin{eqnarray} \label{meson-mass}
M_{\pi^\pm}^2 & = & 2\hat{m}B_0, \hspace{10mm} M_{\pi^0}^2 =
(2\hat{m} - \varepsilon)B_0 + O(\varepsilon^2),\non\\
M_{K^\pm}^2 & = & (m_u+m_s)B_0, \hspace{5mm}
M_{K^0}^2=(m_d+m_s)B_0, \non\\ M_\eta^2 & = & \frac{2}{3}(\hat{m}
+ 2m_s + \frac{3}{2}\varepsilon)B_0 + O(\varepsilon^2),
\end{eqnarray}
where
\begin{equation}
\hat{m}=\frac{1}{2}(m_u+m_d),\hspace{10mm} \varepsilon =
\frac{1}{4}\frac{(m_u-m_d)^2}{(m_s-\hat{m})}.
\end{equation}
Use $B_0=m_\psi\approx 474.13$ Mev and the input
$M_{\pi^\pm}=139.6$ MeV, $M_{K^\pm}=493.677$ MeV,
$M_{K^0}=497.672$ MeV, we obtain that
\begin{eqnarray}
m_u & \approx & 16.37\mbox{ Mev},\hspace{10mm} m_d \approx
24.73\mbox{ Mev}, \hspace{10mm} m_s \approx 497.66\mbox{ Mev}, \non\\
M_{\pi^0}& \approx & 138.97\mbox{ Mev},\hspace{10mm} M_\eta
\approx 566.8\mbox{ Mev}.
\end{eqnarray}
Compared with the experimental values $M_{\pi^0}=134.98$ MeV and
$M_{\eta}=547.45$ MeV. Our results are acceptable. It is noted
that the light quark masses given by particle data group (PDG) are
$m_u = 2\mbox{ to }8$ MeV, $m_d = 5\mbox{ to }15$ MeV, $m_s =
100\mbox{ to }300$ MeV at scale $\mu\approx1$ Gev. It is meaning
that the effective running scale in the order $O(p^2)$ in our
theory should be smaller than 1 Gev. In other words, the suitable
energy region in the order $O(p^2)$ in our theory also should be
smaller than 1 Gev.

Although our result in the order of $O(p^2)$ does coincide with
that in the literature, that in the order of $O(p^4)$ does not.
Our result is
\begin{widetext}
\begin{eqnarray}
{\cal L}_4 & = & -
\frac{f^2}{4m_{\psi}^2}\left\{-\frac{1}{72}\langle
D_{\mu}U^{\dagger}D^{\mu}U\rangle^2 -\frac{1}{36} \langle
D_{\mu}U^{\dagger}D_{\nu}U\rangle\langle
D^{\mu}U^{\dagger}D^{\nu}U\rangle +
\,\frac{1}{12}\langle(D_{\mu}U^{\dagger}D^{\mu}U)^2
\rangle\right.\non \\
& & + \frac{1}{6}\langle F^r_{\mu\nu}F_r^{\mu\nu} +
F^l_{\mu\nu}F_l^{\mu\nu}\rangle + \frac{i}{18}\langle
F^l_{\mu\nu}D^{\mu}U^{\dagger}D^{\nu}U + F^r_{\mu\nu}D^{\mu}U
D^{\nu}U^{\dagger}\rangle-\,\frac{1}{16}\langle U\chi^{\dagger}
- \chi U^{\dagger}\rangle^2\non \\
& & + \frac{5}{8} \langle \chi^{\dagger}\chi \rangle +
\frac{3}{16} \langle \chi^{\dagger}U\chi^{\dagger}U +
U^{\dagger}\chi U^{\dagger}\chi\rangle + \frac{1}{2}\langle
D_{\mu}U^{\dagger}D^{\mu}U(U^{\dagger}\chi + \chi^{\dagger}U)\rangle\non\\
& & +\,\frac{1}{4}\langle \left[\Pi^{\mu},\Delta_{\mu}\right]
E\rangle - \frac{i}{4}\langle\partial_{\mu}\Pi^{\mu}F\rangle -
\frac{4}{9}\langle\left[\Pi_{\mu},\Pi_{\nu}\right]
\left[\Delta^{\mu},\Delta^{\nu}\right]\rangle -
\frac{2}{9}\langle\left[\Pi_{\mu},\Pi_{\nu}\right]^2\rangle\non\\
& & +\,\frac{i}{9}\langle\xi^{\dagger}F^r_{\mu\nu}\xi
\left[2\Delta^{\mu}+\Pi^{\mu},\Pi^{\nu}\right] - \xi F^l_{\mu\nu}
\xi^{\dagger}\left[2\Delta^{\mu}-\Pi^{\mu},\Pi^{\nu}\right]\rangle\non\\
& &\left.-\langle[\frac{2}{9}(\Pi_{\nu}\Delta_{\mu} +
\Delta_{\nu}\Pi_{\mu}) + \frac{10}{9}(\Pi_{\mu}\Delta_{\nu} +
\Delta_{\mu}\Pi_{\nu})] \left[\Pi^{\mu},\Delta^{\nu}\right]
\rangle \right\},
\end{eqnarray}
with
\begin{subequations}
\begin{eqnarray}
\Pi_{\mu}&=&\frac{i}{2}\left[\xi^{\dagger},\partial_{\mu}\xi\right]
+ \frac{1}{2}\xi l_{\mu} \xi^{\dagger} + \frac{1}{2}\xi^{\dagger}
r_{\mu}\xi,\\ \non\\
\Delta_{\mu} & = &
\frac{i}{2}\left\{\xi^{\dagger},\partial_{\mu}\xi\right\} -
\frac{1}{2}\xi l_{\mu} \xi^{\dagger} + \frac{1}{2}\xi^{\dagger}
r_{\mu} \xi\non\\ &=&
\frac{i}{2}\xi^{\dagger}D_{\mu}U\xi^{\dagger} \,=\, -
\frac{i}{2}\xi D_{\mu}U^{\dagger}\xi,\\ \non
\end{eqnarray}
\begin{equation}
E=\xi\chi^{\dagger}\xi - \xi^{\dagger}\chi\xi^{\dagger},\ \
F=\xi\chi^{\dagger}\xi + \xi^{\dagger}\chi\xi^{\dagger},
\end{equation}
\end{subequations}
while the result in the literature~\cite{Gasser:85a} is
\begin{eqnarray}
{\cal L}_4 & = & L_1\langle D_{\mu}U^{\dagger}D^{\mu}U\rangle^2 +
L_2\langle D_{\mu}U^{\dagger}D_{\nu}U\rangle\langle
D^{\mu}U^{\dagger}D^{\nu}U\rangle\non\\
& & +\,L_3\langle D_{\mu}U^{\dagger}D^{\mu}U
D_{\nu}U^{\dagger}D^{\nu}U\rangle + L_4\langle
D_{\mu}U^{\dagger}D^{\mu}U\rangle\langle U^{\dagger}\chi +
\chi^{\dagger}U\rangle\non\\
& & +\,L_5\langle D_{\mu}U^{\dagger}D^{\mu}U\left(U^{\dagger}\chi
+ \chi^{\dagger}U\right)\rangle + L_6\langle U^{\dagger}\chi +
\chi^{\dagger}U\rangle^2\non\\
& & +\,L_7\langle U^{\dagger}\chi - \chi^{\dagger}U\rangle^2 +
L_8\langle \chi^{\dagger}U\chi^{\dagger}U+U^{\dagger}\chi
U^{\dagger}\chi\rangle\non\\
& & -\,iL_9\langle F^l_{\mu\nu}D^{\mu}U^{\dagger}D^{\nu}U +
F^r_{\mu\nu}D^{\mu}U D^{\nu}U^{\dagger}\rangle + L_{10}\langle
U^{\dagger}F^{\mu\nu}_r U F_{l\mu\nu}\rangle\non\\
& & +\,H_1\langle F^r_{\mu\nu}F_r^{\mu\nu} +
F^l_{\mu\nu}F_l^{\mu\nu}\rangle +
H_2\langle\chi^{\dagger}\chi\rangle.
 \label{p^4result}
\end{eqnarray}
\end{widetext}
Our result is not so compact and can not be written in the
covariant form as in the literature. This is because that our
starting point, the CQM Lagrangian in eq.(\ref{CQM-L}), is too
simplistic. In above, in order to obtain the CQM Lagrangian we
made so many simplification. In eq.(\ref{Apropagator}) we absorbed
all the nonperturbative effect to the effective gluon mass. This
procedure is very rough. In eq.(\ref{LS-type}) we ignored the
cubic and other higher order terms. The former is an explicit
chiral symmetry breaking term which should have contribution to
the meson mass. And the latter have something to do with high
energy effect. In ChPT, if the running energy is such high that
have to take the $O(p^4)$ contribution into account then the high
order contribution in eq.(\ref{LS-type}) must also be considered.

After counting the $O(p^4)$ contribution in we can obtain that
\begin{equation}\label{f(p4)}
f_{\pi^\pm} = f-2g_\phi\hat{m}
\end{equation}
and
\begin{equation}\label{M(p4)}
M_\phi^2 = b_\Phi/a_\Phi
\end{equation}
for $\Phi = \pi^0$, $\pi^\pm$, $K^0$, $K^\pm$, $\eta$ with
\begin{eqnarray}\label{aPhi&bPhi}
a_{\pi^0} &=& a_{\pi^\pm} = 1-\frac{2}{B_0}\hat{m},\non\\
a_{K^\pm} &=& 1-\frac{1}{B_0}(m_u+m_s),\non\\
a_{K^0} &=& 1-\frac{1}{B_0}(m_d+m_s),\non\\
a_{\eta} &=& 1-\frac{2}{3B_0}(\hat{m}+2m_s),\non\\
b_{K^0} &=& B_0(m_d+m_s)-\frac{3}{4}(m_d+m_s)^2,\non\\
b_{K^\pm} &=& B_0(m_u+m_s)-\frac{3}{4}(m_u+m_s)^2,\non\\
b_{\pi^\pm} &=& 2B_0\hat{m}-3\hat{m}^2,\non\\
b_{\pi^0} &=& 2B_0\hat{m}-(B_0-2\hat{m}-m_s)\varepsilon -
(4\hat{m}^2-m_um_d)+O(\varepsilon^2),\non\\
b_{\eta} &=& \frac{2}{3B_0}(\hat{m} + 2m_s) -
\frac{1}{3}(4\hat{m}^2 + 4m_s^2 - m_um_d + 4m_s\hat{m}) +
(B_0-2\hat{m}-m_s)\varepsilon + O(\varepsilon^2).
\end{eqnarray}
Same as above, by the input value $f_{\pi^+}\simeq 130.7$ MeV,
$M_{\pi^\pm}=139.6$ MeV, $M_{K^\pm}=493.677$ MeV,
$M_{K^0}=497.672$ MeV, we have
\begin{eqnarray} \label{p4result}
m_u \approx 17.36\mbox{ Mev}, & m_d \approx
20.16\mbox{ Mev}, & m_s \approx 311.45\mbox{ Mev}, \non\\
M_{\pi^0} \approx 134.32\mbox{ Mev}, & M_\eta \approx 283.79\mbox{
Mev}, & f \approx 140.34\mbox{ Mev}.
\end{eqnarray}
It is noted that in eq.(\ref{f(p4)}) and eq.(\ref{aPhi&bPhi}) we
didn't count the loop contribution in. This is because that in our
method all the loop effect should have been absorbed into the LECs
(see Sec.~\ref{Sec.2}), there is no need to consider their
contribution. We will give more discussion about this in later. In
eq.(\ref{p4result}) we can find that our $f$ is indeed vary with
order. This is different from
others'~\cite{Gasser:84a,Espriu:90a}.

In above we ever said that by our method there is no need to
consider loop contribution of meson field or, in other words, to
renormalize the Lagrangian\footnotemark[5]\footnotetext[5]{The
other fields such as weak or lepton particles still need to be
renormalized.}. Most of authors think that renormalizable is
necessary condition when construct an effective field theory. But
we have different opinion about this. We think that there is no
need to renormalize an effective field theory. Why we think so?
Let's recall that why have an effective field theory? Why we need
an effective Lagrangian with new DOFs to cope with problems. In
Sec.~\ref{Sec.1} and \ref{Sec.2} we have said that the coupling
constant in original Lagrangian is too large to carry out
perturbative expansion method. So it is necessary to have an
adequate and effective Lagrangian to describe the system. And the
divergent effect of high order terms in underlying Lagrangian
should be replaced by all sorts of effective terms in the
effective Lagrangian. This idea can be demonstrated as follows,
\begin{eqnarray} \label{toyrelation}
\exp{i\int{\cal L}_0d^4x} & = & 1 + i\int{\cal L}_0d^4x +
\frac{1}{2!}
\left(i\int{\cal L}_0d^4x\right)^2 + \cdots\non\\
& = & 1 + i\int{\cal L_{\rm eff}}d^4x \longrightarrow
\exp{i\int{\cal L_{\rm eff}}d^4x}.
\end{eqnarray}
From eq.(\ref{toyrelation}) we can see that the higher order
contribution of effective action doesn't come from the series
expansion terms but from the higher order terms of the effective
Lagrangian. In fact all the divergent quantities obtained from the
original Lagrangian ${\cal L}_0$ have been renormalized and
absorbed into the LECs in the effective Lagrangian. This
characteristic is very conspicuous in our method.

\section{Conclusion} \label{Sec.5}
It is about twenty years since the ChPT cames into being. Although
it does obtain very good results in low energy physics, still it
can not be treated as a complete theory. There are still some
problems should be considered. In above context, we have shown
that how to solve these problems. We have described a simple
procedure to derive the low-energy effective Lagrangian of the
strong interactions between the octet of pseudoscalar states from
QCD. We see that the complete procedure is straightforward with an
obvious physical concept of transition process. Firstly, by
expanding the action of gluon fields in power of the quark
current, the NJL-type Lagrangian can be obtained easily, which is
that most of theoretical derivative techniques are started in.
Then, by mesonizing this Lagrangian, we see how the constituent
quarks were generated and how the strong interaction effect was
represented in the background meson field. This step also shows
how the dynamical masses of constituent quarks were generated.
Finally, we integrated out the constituent quark DOF and use
derivative expansion method to expand the fermionic determinant.
In this process we find that if the meson particles have positive
mass (for $B_0>0$) then the constituent quarks must has negative
mass. We know that a particle will be a virtual particle if its
mass is negative. And a virtual particle is invisible. Then our
result can just explain why the quarks are not observed in
natural.

Finally, we mention that the Gasser and Leutwyler's ChPT is still
a very successful method when deal with the low energy physics.
Although we think that there is no need to take into account the
loop effect. By their method, due to that the LECs are obtained by
fit in with the experimental values. They can obtain a very good
result even though they count the loop contribution in. They
consider the loop effect of each order in effective Lagrangian and
absorbed it into the LECs. We absorbed the loop effect to the
parameter $f$ which contained in the LECs when construct the
effective Lagrangian. In some respects these two concepts are the
same.

Our model is still roughly. In eq.(\ref{ssbpotential}), in order
to obtain a potential with spontaneous symmetry breaking property
we didn't consider the cubic term, but this term should have more
contribution than the quartic term. So our constituent quark form
Lagrangian, eq.(\ref{CQM-L}), and the effective meson Lagrangian
which derived from this potential are oversimplification. In our
calculation the result which contain $O(p^2)$ and $O(p^4)$ is even
worse than the result which only contain $O(p^2)$ contribution. It
is a fatal defect in perturbation theory. Furthermore, The vector
and pseudovector states of meson fields should also be included.
The question of how to translate the diquark part of interaction
to baryon field remains to be investigated.

\appendix
\section{}
This appendix is a summary of the steps of the calculation in
Sect.~\ref{Sec.4}. From eq.(\ref{effaction}) we know the effective
action without the WZW term is
\begin{equation}
\Gamma_{\rm eff}[U] = i\sum_{n=1}^\infty\frac{(-1)^n}{n}
Tr[(i\cov{$\partial$}+m_\psi)^{-1}\cov{$G$}']^n,
\end{equation}
and the propagator for $\psi$ field is
\begin{equation} \label{propagator}
S_{xy} \equiv (i\cov{$\partial$}+m_{\psi})^{-1}_{xy} =
\int\frac{d^4q}{(2\pi)^4} e^{-iq(x-y)}\frac{\cov{$q$} -
m_{\psi}}{q^2 - m^2_{\psi}+i\epsilon}.
\end{equation}
To simplify the calculation we rewrite the connection $\cov{$G$}'$
as
\begin{equation} \label{connection}
\cov{$G$}'=\gamma^{\mu}\Pi_{\mu} +
\gamma^{\mu}\gamma_5\Delta_{\mu} - \frac{1}{4B_0}\gamma_5E -
\frac{1}{4B_0}F,
\end{equation}
then the first-order term of the effective action is
\begin{eqnarray}
\Gamma^{(1)} & = & -i Tr\int d^4x S_{xx}\cov{$G$}'_x \non\\
& = & -i N_cI_0\frac{m_{\psi}}{B_0} \int d^4x \langle
U^{\dagger}\chi + \chi^{\dagger}U\rangle.
\end{eqnarray}
This term belong to the contribution of $O(p^2)$. Similarly, the
rest can be deduced by analogy:
\begin{equation}
\Gamma^{(2)} = \frac{i}{2} Tr\int d^4x d^4y S_{xy}\cov{$G$}'_x
S_{yx}\cov{$G$}'_x,
\end{equation}
\begin{equation}
\Gamma^{(3)} = \frac{-i}{3} Tr\int d^4x d^4y d^4z
S_{xy}\cov{$G$}'_x S_{yz}\cov{$G$}'_z S_{zx}\cov{$G$}'_x,
\end{equation}
$$\vdots$$Inserting the eq.(\ref{connection}) into the
expressions for the effective action and retaining terms up to
$O(p^4)$ we obtain
\begin{widetext}
\begin{eqnarray}
{\cal L}_2+{\cal L}_4 & = & \frac{f^2}{4}\langle
D_{\mu}U^{\dagger} D^{\mu}U
+ U^{\dagger}\chi + \chi^{\dagger}U\rangle \non \\
& & - \frac{f^2}{4m_{\psi}^2}\left\{-\frac{1}{72}\langle
D_{\mu}U^{\dagger}D^{\mu}U\rangle^2 -\frac{1}{36} \langle
D_{\mu}U^{\dagger}D_{\nu}U\rangle\langle
D^{\mu}U^{\dagger}D^{\nu}U\rangle
+\,\frac{1}{12}\langle(D_{\mu}U^{\dagger}D^{\mu}U)^2
\rangle\right.\non \\
& & + \frac{1}{6}\langle F^r_{\mu\nu}F_r^{\mu\nu} +
F^l_{\mu\nu}F_l^{\mu\nu}\rangle + \frac{i}{18}\langle
F^l_{\mu\nu}D^{\mu}U^{\dagger}D^{\nu}U + F^r_{\mu\nu}D^{\mu}U
D^{\nu}U^{\dagger}\rangle-\,\frac{1}{16}\langle U\chi^{\dagger}
- \chi U^{\dagger}\rangle^2\non \\
& & + \frac{5}{8} \langle \chi^{\dagger}\chi \rangle +
\frac{3}{16} \langle
\chi^{\dagger}U\chi^{\dagger}U+U^{\dagger}\chi
U^{\dagger}\chi\rangle + \frac{1}{2}\langle
D_{\mu}U^{\dagger}D^{\mu}U(U^{\dagger}\chi + \chi^{\dagger}U)\rangle\non\\
& & +\,\frac{1}{4}\langle \left[\Pi^{\mu},\Delta_{\mu}\right]
E\rangle - \frac{i}{4}\langle\partial_{\mu}\Pi^{\mu}F\rangle -
\frac{4}{9}\langle\left[\Pi_{\mu},\Pi_{\nu}\right]
\left[\Delta^{\mu},\Delta^{\nu}\right]\rangle -
\frac{2}{9}\langle\left[\Pi_{\mu},\Pi_{\nu}\right]^2\rangle\non\\
& & +\,\frac{i}{9}\langle\xi^{\dagger}F^r_{\mu\nu}\xi
\left[2\Delta^{\mu}+\Pi^{\mu},\Pi^{\nu}\right] - \xi F^l_{\mu\nu}
\xi^{\dagger}\left[2\Delta^{\mu}-\Pi^{\mu},\Pi^{\nu}\right]\rangle\non\\
& &\left.-\langle[\frac{2}{9}(\Pi_{\nu}\Delta_{\mu} +
\Delta_{\nu}\Pi_{\mu}) + \frac{10}{9}(\Pi_{\mu}\Delta_{\nu} +
\Delta_{\mu}\Pi_{\nu})] \left[\Pi^{\mu},\Delta^{\nu}\right]
\rangle \right\},
\end{eqnarray}
where we have used the following identities for simplification:
\begin{equation}
\left[\Delta_{\mu},\Delta_{\nu}\right] = \frac{1}{4}\xi
\left(D_{\mu}U^{\dagger}D_{\nu}U - D_{\nu}U^{\dagger}D_{\mu}U
\right)\xi^{\dagger},
\end{equation}
\begin{equation}
\Pi_{\mu\nu} \,\equiv\, \partial_{\mu}\Pi_{\nu} -
\partial_{\nu}\Pi_{\mu} -i\left[\Pi_{\mu},\Pi_{\nu}\right]
\,=\, \frac{1}{2}\xi^{\dagger}F^r_{\mu\nu}\xi + \frac{1}{2}\xi
F^l_{\mu\nu}\xi^{\dagger}+i\left[\Delta_{\mu},\Delta_{\nu}\right],
\end{equation}
\begin{equation}
\left(D_{\mu}D^{\mu}U\right)U^{\dagger} - U\left(D_{\mu}
D^{\mu}U^{\dagger}\right) = \frac{1}{3}\langle U\chi^{\dagger}
-\chi U^{\dagger}\rangle - \left(U\chi^{\dagger} - \chi
U^{\dagger}\right),
\end{equation}
\begin{equation}
F^l_{\mu\nu}=\partial_{\mu}l_{\nu}-\partial_{\nu}l_{\mu} -
i\left[l_{\mu},l_{\nu}\right],\ \ \ \ F^r_{\mu\nu} =
\partial_{\mu}r_{\nu}-\partial_{\nu}r_{\mu} -
i\left[r_{\mu},r_{\nu}\right],
\end{equation}
\begin{equation}
\left(D_{\mu}D_{\nu}-D_{\nu}D_{\mu}\right)U = iUF^l_{\mu\nu} -
iF^r_{\mu\nu}U,\ \ \
\left(D_{\mu}D_{\nu}-D_{\nu}D_{\mu}\right)U^{\dagger} =
iU^{\dagger}F^r_{\mu\nu} - iF^l_{\mu\nu}U^{\dagger}.
\end{equation}
\end{widetext}

\end{document}